\documentclass[aps]{revtex4}
\usepackage{amsmath}
\usepackage{graphicx}

\begin{document}

\title{Interference induced preparation of spinpolarized electrons in a three-terminal quantum ring}
\author{O K\'{a}lm\'{a}n$^{1,2}$, P F\"{o}ldi$^1$, M G Benedict$^1$ and F M
Peeters$^2$}
\address{$^1$Department of Theoretical Physics, University of Szeged, Tisza
Lajos k\"{o}r\'{u}t 84-86, H-6720 Szeged, Hungary} 
\address{$^2$Departement Fysica, Universiteit Antwerpen, Groenenborgerlaan 171, B-2020
Antwerpen, Belgium}

\begin{abstract}
We present an exact, analytic solution of the spin dependent quantum
transport problem with spin-orbit interaction in a one-dimensional
mesoscopic ring with one input and two output leads. We demonstrate that for
appropriate parameters spatial interference in the ring leads to a behavior
analogous to that of the Stern-Gerlach apparatus: different spin
polarizations can be achieved in the two output channels from an originally
totally unpolarized incoming spin state. It is shown that this requires an
appropriate interference of states that carry oppositely directed currents.
We find that spin polarization is possible for several geometries, including
the case when the device is not symmetric with respect to the incoming lead.
A clear connection is established between the Stern-Gerlach like property of
the device and the relevant Aharonov-Casher phases in the loop geometry.
\end{abstract}

\maketitle

\section{Introduction}

Electronics based on the spin of the electron represents a new direction of
development (spintronics) \cite{ZFS04}. In order to utilize this additional
degree of freedom as an (either classical or possibly quantum) computational
resource it is required to develop a controllable way of manipulating spins.
One of the most promising mechanisms that can be used for this purpose is
the spin-orbit interaction \cite{R60} in semiconductor materials, which can
be controlled by an external electric field \cite{NATE97}. For semiconductor
nanostructures, where the mean free path of the electron can be much larger
than the size of the device, quantum mechanical interference can lead to a
new class of spin-sensitive devices, such as quantum gates \cite{FMBP05}. On
the other hand, even if a whole spintronic apparatus does not use the
quantum mechanical nature of the electron for information processing, some
parts of it still may rely on interference phenomena, similarly to the
polarizing device discussed in this paper.

Quantum rings \cite{AL93,NMT99,BIA84,VKPB06} or loops \cite{KNV04} (i.e.,
ring shaped objects where quantum mechanical interference plays an important
role) made of a semiconductor material, have been shown to have remarkable
spin transformation properties \cite%
{NATE97,NMT99,SKY01,MPV04,KNV04,FR04,FMBP05,ZX05,ID03,G02}. This is
partially due to the geometry of these devices, as the incoming electrons
are forced to split into two different spatial parts that interfere at the
output, while the spin-sensitive interaction introduces an additional effect
to be taken into account. Consequences of the interplay between spatial
interference and the so-called Rashba-type \cite{R60} spin-orbit interaction
was recently observed in HgTe nanorings, where an external magnetic field
was also present \cite{KTHS06}. More generally, the spin degree of freedom
in quantum interference \cite{SN05,KMGA05,CCZ04,PPC06} can play a prominent
role in the development of a spintronic network based on various
spin-sensitive devices \cite{EBL03,SB03,YPS02,FHR01}.

Let us recall that already Bohr and Mott pointed out \cite{MM49}, that in
contrast to atoms, one can not spin-polarize electrons in an inhomogeneous
magnetic field. We consider here a three-terminal quantum ring, where
electrons entering in a totally unpolarized spin state become polarized at
the outputs with different spin directions. This device can be deemed in a
certain sense a spintronic analogue of the Stern-Gerlach apparatus \cite%
{FKBP06}. A related polarizing effect was recently predicted in a Y-shaped
conductor which was a consequence of scattering on impurities \cite{P04}.
Note that this is a very different physical mechanism from the coherent spin
transfer to be discussed here. Our model is based on an exact, analytic
solution of the spin dependent transport problem and thus provides a clear
physical picture of a process where fundamental polarization effects as well
as nontrivial spatial-spin correlations, entanglement or intertwining \cite%
{note1} appear \cite{KFB06}. In addition, our treatment allows us to
determine the parameter values for which the device is reflectionless, i.e.
perfect polarization at the outputs takes place without losses.

In the present paper we demonstrate that the physical origin of the
polarizing effect we found earlier in Ref.~\cite{FKBP06} is essentially
spatial interference: At a certain output junction the spatial parts of one
of the eigenspinors representing clockwise and anticlockwise directed
currents interfere destructively, leading to the transmission of the other
orthogonal eigenspinor in the output lead. This effect is visualized by
plotting the spatial dependence of the spin direction along the ring. We
show that there are several, not necessarily symmetric positions where such
destructive interference takes place. All of these configurations possess
the polarizing property. We also determine the conditions for the spin
polarization effect in terms of the Aharonov-Casher phases \cite{AC84}
gained by the two orthogonal eigenspinors.

\bigskip

\section{The model of spin dependent scattering in a ring}

The one-dimensional Hamiltonian of an electron moving on a ring situated in
the $x-y$ plane in the presence of Rashba spin-orbit interaction is given by%
\cite{MPV04,MMK02}%
\begin{equation}
H=\hbar \Omega \left[ \left( -i\frac{\partial }{\partial \varphi }+\frac{%
\omega }{2\Omega }(\sigma _{x}\cos \varphi +\sigma _{y}\sin \varphi )\right)
^{2}-\frac{\omega ^{2}}{4\Omega ^{2}}\right] ,  \label{Ham}
\end{equation}%
where $\varphi $ is the azimuthal angle of a point on the ring, $\hbar
\Omega =\hbar ^{2}/2m^{\ast }a^{2}$ is the dimensionless kinetic energy of
the electron, $a$ is the radius of the ring, $m^{\ast }$ denotes the
effective mass of the electron, and $\omega =e\hbar E_{z}/\left( \sqrt{2}%
m^{\ast }ac\right) ^{2}$ is the frequency associated with the spin-orbit
interaction, with $E_{z}$\ being a static electric field perpendicular to
the surface of the ring. The parameter $\omega $ can be tuned with an
external gate voltage \cite{NATE97}. Apart from constants, this Hamiltonian (%
\ref{Ham}) is the square of the sum of the $z$ component of the orbital
angular momentum operator $L_{z}=-i\frac{\partial }{\partial \varphi }$,
and of $\frac{\omega }{\Omega }S_{r}$, where $S_{r}=\sigma _{r}/2$ is the
radial component of the spin (both measured in units of $\hbar )$. $H$
commutes with the $z$ component of the total angular momentum $K=L_{z}+S_{z}$%
, as well as with $S_{\theta \varphi }=S_{x}\sin \theta \cos \varphi
+S_{y}\sin \theta \sin \varphi +S_{z}\cos \theta $, the spin component in
the direction determined by the angles $\theta $, and $\varphi $, where $%
\theta $ is given by $\tan \theta =-\omega /\Omega $. Therefore a basis can
be constructed which are simultaneous eigenfunctions of the operators $H$, $%
K $, and $S_{\theta \varphi }$. In the $\left\vert \uparrow \right\rangle $, 
$\left\vert \downarrow \right\rangle $ eigenbasis of $S_{z}$ we can find
these in the form \cite{FMBP05}:%
\begin{equation}
\psi (\kappa ,\varphi )=e^{i\kappa \varphi }%
\begin{pmatrix}
e^{-i\varphi /2}u(\kappa ) \\ 
e^{i\varphi /2}v(\kappa )%
\end{pmatrix}%
,  \label{est}
\end{equation}%
obeying 
\begin{eqnarray*}
K\psi \left( \kappa ,\varphi \right) &=&\kappa \psi \left( \kappa ,\varphi
\right) , \\
S_{\theta \varphi }\psi \left( \kappa ,\varphi \right) &=&\pm \frac{1}{2}%
\psi \left( \kappa ,\varphi \right) .
\end{eqnarray*}%
The energy eigenvalues are%
\begin{eqnarray}
E_{\mu } &=&\hbar \Omega \left[ \left( \kappa -\frac{1}{2}-\frac{\Phi
^{\left( \mu \right) }}{2\pi }\right) ^{2}-\frac{\omega ^{2}}{4\Omega ^{2}}%
\right]  \notag \\
&=&\hbar \Omega \left[ \kappa ^{2}+\left( -1\right) ^{\mu }\kappa w+1/4%
\right] ,  \label{En}
\end{eqnarray}%
where $\mu =1,2$, $w=\sqrt{1+\omega ^{2}/\Omega ^{2}}$ and $\Phi ^{\left(
\mu \right) }=-\pi \left[ 1+\left( -1\right) ^{\mu }w\right] $ is the
Aharonov-Casher phase \cite{AC84}. In a ring connected to leads the energy
is a continuous variable -- since $\left( \kappa \pm 1/2\right) $ is no
longer an integer as it is in the case of a closed ring -- and the possible
values of $\kappa $ are the solutions of equation (\ref{En}) for $E_{\mu }=E$%
: 
\begin{eqnarray}
\kappa _{j}^{\mu } &=&\frac{1}{2}+\frac{\Phi ^{\left( \mu \right) }}{2\pi }%
+\left( -1\right) ^{\mu +j+1}q  \notag \\
&=&\left( -1\right) ^{\mu +1}(w/2+(-1)^{j}q),  \label{kappa_j_mu}
\end{eqnarray}%
where $\mu ,j=1,2$ and $q=\sqrt{(\omega /2\Omega )^{2}+E/\hbar \Omega }$.
The energy eigenvalues of the Hamiltonian are fourfold degenerate, thus the
state of the electron in the different segments of the ring (see figure \ref%
{fig1}) for a given $E$ is a linear combination of the corresponding four $%
\psi (\kappa _{j}^{\mu },\varphi )$\ eigenstates%
\begin{equation}
\Psi _{i}(\varphi )=\sum_{\mu ,j=1,2}a_{ij}^{\mu }\psi (\kappa _{j}^{\mu
},\varphi ),\qquad i=I,I\!I,I\!I\!I,  \label{Psi_i}
\end{equation}%
where the ratio of the components of the spinors in (\ref{est}) is given by%
\begin{equation}
\frac{v_{j}^{\mu }}{u_{j}^{\mu }}=\frac{v(\kappa _{j}^{\mu })}{u(\kappa
_{j}^{\mu })}=\tan {\theta }^{\left( \mu \right) }{/2}=\frac{\Omega }{\omega 
}\left( 1+\left( -1\right) ^{\mu }w\right) .  \label{theta}
\end{equation}%
Since $\tan {\theta }^{\left( 1\right) }{/2}=-\cot {\theta }^{\left(
2\right) }{/2}$, we can express the two eigenspinors with $\theta ^{\left(
1\right) }=\theta $:%
\begin{equation}
\psi _{j}^{1}(\kappa _{j}^{1},\varphi )=e^{i\kappa _{j}^{1}\varphi }%
\begin{pmatrix}
e^{-i\varphi /2}\cos \frac{\theta }{2} \\ 
e^{i\varphi /2}\sin \frac{\theta }{2}%
\end{pmatrix}%
,  \label{u1v1}
\end{equation}%
\begin{equation}
\psi _{j}^{2}(\kappa _{j}^{2},\varphi )=e^{i\kappa _{j}^{2}\varphi }%
\begin{pmatrix}
e^{-i\varphi /2}\sin \frac{\theta }{2} \\ 
-e^{i\varphi /2}\cos \frac{\theta }{2}%
\end{pmatrix}%
.  \label{u2v2}
\end{equation}

\begin{figure}[tbp]
\begin{center}
\includegraphics*[width=12 cm]{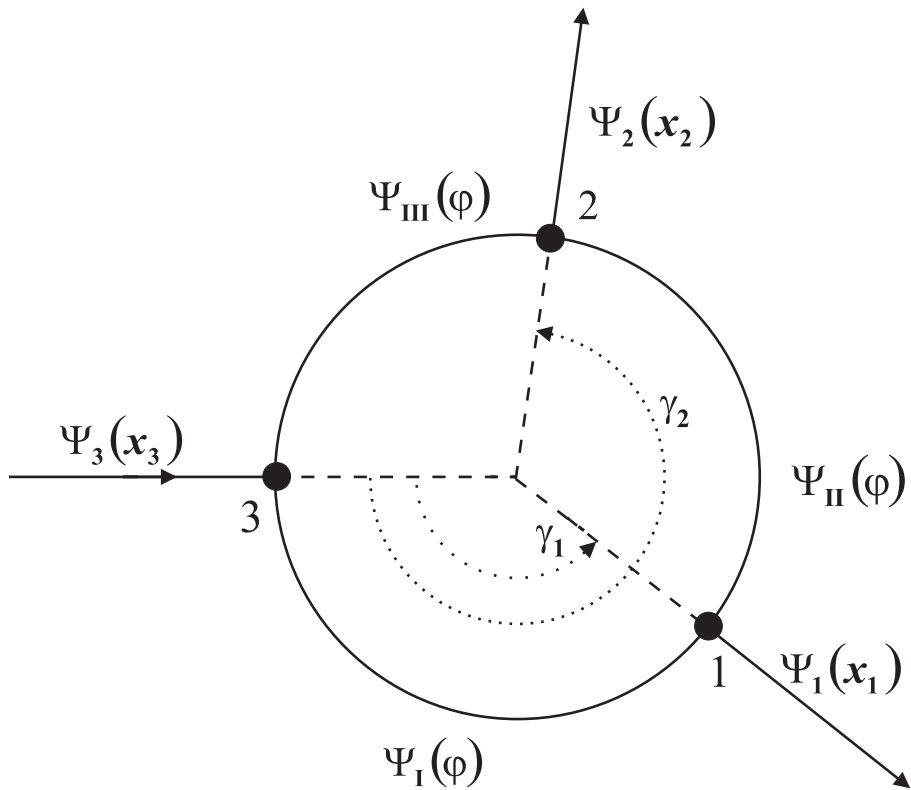}
\end{center}
\caption{The geometry of the device and the relevant wave functions in the
different domains. The parameter $\protect\varphi $ is measured from
junction \textbf{3} in counterclockwise direction.}
\label{fig1}
\end{figure}

The stationary states of the complete problem (ring and leads), can be
determined by imposing continuity of the wave functions at the boundary of
the different domains. Using local coordinates as shown in figure \ref{fig1}%
, the incoming wave, $\Psi _{3}(x_{3})$, and the outgoing waves $\Psi
_{1}(x_{1}),$ $\Psi _{2}(x_{2})$ are built up of linear combinations of
spinors with spatial dependence $e^{ikx}$ etc. corresponding to $E={\hbar
^{2}k^{2}}/{2m^{\ast }}$: 
\begin{eqnarray}
\Psi _{3}\left( x_{3}\right) &=&%
\begin{pmatrix}
f_{\uparrow } \\ 
f_{\downarrow }%
\end{pmatrix}%
e^{i kx_{3}}+%
\begin{pmatrix}
r_{\uparrow } \\ 
r_{\downarrow }%
\end{pmatrix}%
e^{-i kx_{3}},  \notag \\
\Psi _{n}\left( x_{n}\right) &=&%
\begin{pmatrix}
t_{\uparrow }^{n} \\ 
t_{\downarrow }^{n}%
\end{pmatrix}%
e^{i kx_{n}},  \label{waves}
\end{eqnarray}%
where $n=1,2.$ Our aim is to determine the transmission properties of the
ring, i.e. to find the elements of the transmission matrices, which are
defined in the following way%
\begin{equation}
T^{\left( n\right) }%
\begin{pmatrix}
f_{\uparrow } \\ 
f_{\downarrow }%
\end{pmatrix}%
=%
\begin{pmatrix}
t_{\uparrow }^{n} \\ 
t_{\downarrow }^{n}%
\end{pmatrix}%
,  \label{T}
\end{equation}

where $n=1,2$ label the two outgoing leads.

In order to obtain the transmission matrices, the 12 coefficients $%
a_{ij}^{\mu }$ have to be determined. This can be done via requiring
continuity of the wave functions, and vanishing net spin current densities
(Griffith conditions) \cite{G53,X92,MPV04,FMBP05} at the three junctions.

According to the detailed calculations presented in the Appendix, the
elements of the transmission matrices are:%
\begin{eqnarray}
T_{\uparrow \uparrow }^{\left( n\right) } &=&\frac{8qka}{y}e^{-i\frac{%
\gamma n}{2}}\left[ \cos ^{2}\frac{\theta }{2}\left( h_{1}^{\left( n\right)
}+h_{2}^{\left( n\right) }\right) +\sin ^{2}\frac{\theta }{2}\left(
h_{1}^{\left( n\right) }-h_{2}^{\left( n\right) }\right) ^{\ast }\right] , 
\notag \\
T_{\uparrow \downarrow }^{\left( n\right) } &=&\frac{8qka}{y}e^{-i%
\frac{\gamma _{n}}{2}}\sin \frac{\theta }{2}\cos \frac{\theta }{2}\left[
\left( h_{1}^{\left( n\right) }+h_{2}^{\left( n\right) }\right) -\left(
h_{1}^{\left( n\right) }-h_{2}^{\left( n\right) }\right) ^{\ast }\right] ,
\label{T_asymm} \\
T_{\downarrow \uparrow }^{\left( n\right) } &=&e^{i\gamma
_{n}}T_{\uparrow \downarrow }^{\left( n\right) },  \notag \\
T_{\downarrow \downarrow }^{\left( n\right) } &=&\frac{8qka}{y}e^{i%
\frac{\gamma _{n}}{2}}\left[ \sin ^{2}\frac{\theta }{2}\left( h_{1}^{\left(
n\right) }+h_{2}^{\left( n\right) }\right) +\cos ^{2}\frac{\theta }{2}\left(
h_{1}^{\left( n\right) }-h_{2}^{\left( n\right) }\right) ^{\ast }\right] , 
\notag
\end{eqnarray}%
where 
\begin{eqnarray*}
h_{1}^{\left( 1\right) } &=&-kae^{i\frac{w}{2}\gamma _{1}}\sin \left(
q\left( 2\pi -\gamma _{2}\right) \right) \sin \left( q\left( \gamma
_{2}-\gamma _{1}\right) \right) , \\
h_{1}^{\left( 2\right) } &=&kae^{i\frac{w}{2}\gamma _{2}}e^{-iw\pi
}\sin \left( q\gamma _{1}\right) \sin \left( q\left( \gamma _{2}-\gamma
_{1}\right) \right) , \\
h_{2}^{\left( n\right) } &=&iqe^{i\frac{w}{2}\gamma _{n}}\left( e%
^{-iw\pi }\sin \left( q\gamma _{n}\right) -\sin \left( q\left( 2\pi -\gamma
_{n}\right) \right) \right) ,
\end{eqnarray*}%
and $y$ is given by (\ref{y}).

The reflection matrix $R$ is found to be diagonal in
the $\{\left\vert \uparrow \right\rangle $, $\left\vert \downarrow
\right\rangle \}$ basis of $S_{z}$:%
\begin{eqnarray*}
  R_{\uparrow \uparrow } &=&R_{\downarrow \downarrow }=\frac{8ka}{y}%
\left\{ -i q^{2}\sin \left( 2q\pi \right) +i k^{2}a^{2}\sin \left(
q\left( \gamma _{2}-\gamma _{1}\right) \right) \sin \left( q\left( 2\pi
-\gamma _{2}\right) \right) \sin \left( q\gamma _{1}\right) \right.  \\
  &&\left. -qka\left[ \sin \left( q\left( 2\pi -\gamma _{1}\right) \right)
\sin \left( q\gamma _{1}\right) +\sin \left( q\left( 2\pi -\gamma
_{2}\right) \right) \sin \left( q\gamma _{2}\right) \right] \right\} -1, \\
  R_{\uparrow \downarrow } &=&R_{\downarrow \uparrow }=0.
\end{eqnarray*}%
This matrix describes the  losses in the efficiency of the spin transformation, as
the sum of the norms of the outgoing and reflected waves should be equal to the 
norm of the input. 

\section{Analysis and visualization}

When the incoming electron is not perfectly spin-polarized, i.e. its spin
state is a mixture, which -- instead of a two component spinor -- should be
described by a $2\times 2$ density matrix $\varrho _{in}$, then we can
readily generalize equation (\ref{T}) to obtain 
\begin{equation}
\nonumber
\varrho ^{\left( n\right) }=T^{\left( n\right) }\varrho _{in}\left(
T^{\left( n\right) }\right) ^{\dagger },
\end{equation}%
where $\varrho ^{\left( 1\right) }$ and $\varrho ^{\left( 2\right) }$ are
the output density matrices in the respective leads.

Considering a completely unpolarized input, i.e. $\varrho _{in}$ being
proportional to the $2\times 2$\ identity matrix, in order to get polarized
outputs, the relevant output density operators should be projectors (apart
from the possible reflective losses): 
\begin{equation}
\frac{1}{2}T^{\left( n\right) }\left( T^{\left( n\right) }\right) ^{\dagger
}=\eta _{n}\left\vert \phi ^{n}\right\rangle \left\langle \phi
^{n}\right\vert .  \label{proj}
\end{equation}%
The non-negative numbers $\eta _{1}$ and $\eta _{2}$ measure the efficiency
of the polarizing device, i.e. $\eta _{1}+\eta _{2}=1$ means a
reflectionless process. Equation (\ref{proj}) is equivalent to require that
the determinants of $T^{\left( n\right) }\left( T^{\left( n\right) }\right)
^{\dagger }$ vanish. According to equations (\ref{T_asymm}) there are two
different conditions for each transmission matrix to satisfy this
requirement: 
\begin{equation}
h_{1}^{\left( n\right) }\pm h_{2}^{\left( n\right) }=0,  \label{detconds}
\end{equation}%
where $n=1,2$ indicates the two output juctions. It can be shown that only
the following two cases lead to nonzero transmission at both outputs: 
\begin{subequations}
\begin{eqnarray}
h_{1}^{\left( 1\right) }+h_{2}^{\left( 1\right) } &=&0,  \notag \\
h_{1}^{\left( 2\right) }-h_{2}^{\left( 2\right) } &=&0,  \label{detcond1}
\end{eqnarray}%
or 
\begin{eqnarray}
h_{1}^{\left( 1\right) }-h_{2}^{\left( 1\right) } &=&0,  \notag \\
h_{1}^{\left( 2\right) }+h_{2}^{\left( 2\right) } &=&0.  \label{detcond2}
\end{eqnarray}%
After substitution we obtain two equations for $\cos \left( w\pi \right) $
and $\sin \left( w\pi \right) $ in both cases: 
\end{subequations}
\begin{subequations}
\label{pol_asymm}
\begin{eqnarray}
 \cos \left( w\pi \right) \! &=&\!\frac{\sin \left( q\left( 2\pi -\gamma
_{1}\right) \right) }{\sin \left( q\gamma _{1}\right) }\!=\!\frac{\sin
\left( q\gamma _{2}\right) }{\sin \left( q\left( 2\pi -\gamma _{2}\right)
\right) },  \label{asymm_cos} \\
 \sin \left( w\pi \right) \! &=&\!\pm \frac{ka}{q}\frac{\sin \left(
q\left( 2\pi -\gamma _{2}\right) \right) \sin \left( q\left( \gamma
_{2}-\gamma _{1}\right) \right) }{\sin \left( q\gamma _{1}\right) }\!=\!\pm 
\frac{ka}{q}\frac{\sin \left( q\gamma _{1}\right) \sin \left( q\left( \gamma
_{2}-\gamma _{1}\right) \right) }{\sin \left( q\left( 2\pi -\gamma
_{2}\right) \right) },  \label{asymm_sin}
\end{eqnarray}%
or with the Aharonov-Casher phase 
\end{subequations}
\begin{subequations}
\begin{eqnarray}
\cos \Phi ^{\left( \mu \right) }\! &=&\!-\!\frac{\sin \left( q\left( 2\pi
-\gamma _{1}\right) \right) }{\sin \left( q\gamma _{1}\right) }\!=\!-\!\frac{%
\sin \left( q\gamma _{2}\right) }{\sin \left( q\left( 2\pi -\gamma
_{2}\right) \right) }, \\
\sin \Phi ^{\left( \mu \right) }\! &=&\!\pm \left( -1\right) ^{\mu }\frac{ka%
}{q}\frac{\sin \left( q\left( 2\pi -\gamma _{2}\right) \right) \sin \left(
q\left( \gamma _{2}-\gamma _{1}\right) \right) }{\sin \left( q\gamma
_{1}\right) }\!  \notag \\
&=&\!\pm \left( -1\right) ^{\mu }\frac{ka}{q}\frac{\sin \left( q\gamma
_{1}\right) \sin \left( q\left( \gamma _{2}-\gamma _{1}\right) \right) }{%
\sin \left( q\left( 2\pi -\gamma _{2}\right) \right) },
\end{eqnarray}%
where the plus sign in (\ref{asymm_sin}) corresponds to equation (\ref%
{detcond1}), while the minus sign is for the case given by equation (\ref%
{detcond2}). From (\ref{asymm_cos}) and (\ref{asymm_sin}) for both signs we
find 
\end{subequations}
\begin{eqnarray*}
\sin \left( q\gamma _{1}\right) &=&\pm \sin \left( q\left( 2\pi -\gamma
_{2}\right) \right) , \\
\sin \left( q\gamma _{2}\right) &=&\pm \sin \left( q\left( 2\pi -\gamma
_{1}\right) \right) .
\end{eqnarray*}%
The solutions of these equations for a fixed $\gamma _{2}$ are%
\begin{equation}
\gamma _{1}=2\pi -\gamma _{2}\pm m\pi /q,  \label{gamma1}
\end{equation}%
or for a fixed $\gamma _{1}$ are%
\begin{equation}
\gamma _{2}=2\pi -\gamma _{1}\pm l\pi /q,  \label{gamma2}
\end{equation}%
where $m,l$ are nonnegative integers which ensure $0\!\leq \!\gamma
_{1,}\gamma _{2}\!\leq \!2\pi $ and $\gamma _{2}\!\geq \!\gamma _{1}$. The $%
m,l=0$ cases correspond to a ring the outgoing leads of which are symmetric
with respect to the incoming lead. It was demonstrated that for such a ring
one can find lines in the $\{\gamma _{2},\omega /\Omega ,ka\}$ space along
which the conditions (\ref{asymm_cos}) and (\ref{asymm_sin}) can be
satisfied \cite{FKBP06}, i.e. the ring polarizes a completely unpolarized
input. Polarization occurs with equal $\eta _{1}=\eta _{2}\equiv \eta /2$
transmission in both outputs. Parameter combinations, for which the
transmission probability $\eta $\ is unity can also be found.

From equations (\ref{gamma1}) and (\ref{gamma2}) we see that we can extend
the polarizing property to asymmetric geometries. The asymmetric positions
of the two output leads for which the condition for complete spin
polarization is satisfied are $\pm l\pi /q$ and $\pm m\pi /q$ angles away
from the symmetric ones. For proper combinations of the $\omega /\Omega $
and $ka$ parameters the asymmetric ring also produces polarized outputs with
equal transmission probabilities $\eta _{1}=\eta
_{2}=128q^{2}k^{2}a^{2}\left\vert h_{1}^{\left( 1\right) }\right\vert
^{2}/\left\vert y\right\vert ^{2}$.\ We note that this is an important
generalization of the results of Ref. \cite{FKBP06}. There are several
appropriate positions for the output leads, the symmetric case is just one
of them.

The output spinors are the eigenstates $\left\vert \phi ^{n}\right\rangle $\
of the transmitted density matrices $\frac{1}{2}T^{\left( n\right) }\left(
T^{\left( n\right) }\right) ^{\dagger }$, which correspond to the nonzero
eigenvalues given by $\eta _{n}$. Focusing on the case of equations (\ref%
{detcond1}), these eigenstates read%
\begin{equation}
\left\vert \phi _{a}^{1}\right\rangle =%
\begin{pmatrix}
e^{-i\frac{\gamma _{1}}{2}}\sin \frac{\theta }{2} \\ 
-e^{i\frac{\gamma _{1}}{2}}\cos \frac{\theta }{2}%
\end{pmatrix}%
,\quad \left\vert \phi _{a}^{2}\right\rangle =%
\begin{pmatrix}
e^{-i\frac{\gamma _{2}}{2}}\cos \frac{\theta }{2} \\ 
e^{i\frac{\gamma _{2}}{2}}\sin \frac{\theta }{2}%
\end{pmatrix}%
.  \label{out1}
\end{equation}%
These results describe the connection between the strength of the spin-orbit
coupling (encoded in $\theta $), the geometry of the device and its
polarizing directions. Note that these spinors are in general
non-orthogonal, their overlap is given by $\left\langle \phi ^{2}\right\vert
\left. \phi ^{1}\right\rangle =i\sin \theta \sin {(\gamma _{2}-\gamma
_{1})/2}$. For the other case given by equations (\ref{detcond2}) we have: 
\begin{equation}
\left\vert \phi _{b}^{1}\right\rangle =%
\begin{pmatrix}
e^{-i\frac{\gamma _{1}}{2}}\cos \frac{\theta }{2} \\ 
e^{i\frac{\gamma _{1}}{2}}\sin \frac{\theta }{2}%
\end{pmatrix}%
,\quad \left\vert \phi _{b}^{2}\right\rangle =%
\begin{pmatrix}
e^{-i\frac{\gamma _{2}}{2}}\sin \frac{\theta }{2} \\ 
-e^{i\frac{\gamma _{2}}{2}}\cos \frac{\theta }{2}%
\end{pmatrix}%
.  \label{out2}
\end{equation}%
We can see from (\ref{out1}) and (\ref{out2}) that the output spin states in
both cases are the two eigenspinors of the Hamiltonian at the positions of
the output junctions. For a given output lead, the two eigenspinors are
interchanged in the two cases.

%
\begin{figure}[tbp]
\begin{center}
\includegraphics*[width=12 cm]{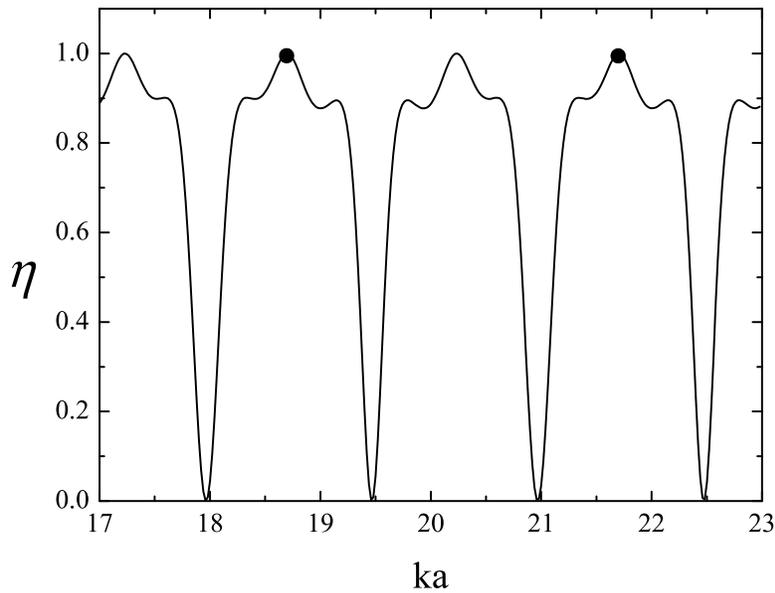}
\end{center}
\caption{Transmission probability $\protect\eta=2\protect\eta _{1}=2\protect%
\eta _{2}$ at the two outputs of an asymmetric ring as a function of $ka$
for $\protect\omega /\Omega =2.27$. This figure corresponds to $\protect%
\gamma _{2}=4\protect\pi /3$ and $\protect\gamma _{1}=2\protect\pi -\protect%
\gamma _{2}-6\protect\pi /q$. The dots mark the points where perfect
polarization occurs.}
\label{fig2}
\end{figure}

Figure \ref{fig2} shows this transmission probability as a function of $ka$
for $\omega /\Omega =2.27$. The dots on the curve mark the points where
perfect polarization occurs. Since the angle $\gamma _{1}$ is a function of $%
ka$, the dots correspond to different asymmetric configurations. It can be
seen that even for an asymmetric ring, with appropriate parameter values,
complete output spin polarization can be achieved with practically zero
reflective loss.

Now we investigate the physical origin of this polarizing effect. To this
end we consider a completely unpolarized input taken as the following equal
weight sum of \textit{pure} state projectors%
\begin{equation}
\nonumber
\varrho _{in}=\frac{1}{2}\left( \left\vert \psi _{in}^{1}\right\rangle
\left\langle \psi _{in}^{1}\right\vert +\left\vert \psi
_{in}^{2}\right\rangle \left\langle \psi _{in}^{2}\right\vert \right) .
\end{equation}%
Here $\psi _{in}^{\mu }=\psi _{1}^{\mu }(0)=\psi _{2}^{\mu }(0)$ ($\mu =1,2$%
),\ given by equations (\ref{u1v1}) and (\ref{u2v2}), are the eigenspinors
of the Hamiltonian at the position of the incoming lead (\textbf{3}). The
density operator in the different sections of the ring is then%
\begin{equation}
\varrho _{i}\left( \varphi \right) =\frac{1}{2}\left( \left\vert \Psi
_{i}^{1}\left( \varphi \right) \right\rangle \left\langle \Psi
_{i}^{1}\left( \varphi \right) \right\vert +\left\vert \Psi _{i}^{2}\left(
\varphi \right) \right\rangle \left\langle \Psi _{i}^{2}\left( \varphi
\right) \right\vert \right) \qquad i=I,I\!I,I\!I\!I,  \label{rho_i}
\end{equation}%
where%
\begin{eqnarray}
\Psi _{i}^{1}\left( \varphi \right) &=&\sum_{j=1,2}\Psi _{ij}^{1}\left(
\varphi \right) =N_{i}^{1}\left( \varphi \right) 
\begin{pmatrix}
e^{-i\frac{\varphi }{2}}\cos \frac{\theta }{2} \\ 
e^{i\frac{\varphi }{2}}\sin \frac{\theta }{2}%
\end{pmatrix}%
,  \notag \\
\Psi _{i}^{2}\left( \varphi \right) &=&\sum_{j=1,2}\Psi _{ij}^{2}\left(
\varphi \right) =N_{i}^{2}\left( \varphi \right) 
\begin{pmatrix}
e^{-i\frac{\varphi }{2}}\sin \frac{\theta }{2} \\ 
-e^{i\frac{\varphi }{2}}\cos \frac{\theta }{2}%
\end{pmatrix}%
,  \label{Psi_i_1_2}
\end{eqnarray}%
are the spinor valued wave functions of the electron in the different
domains of the ring for the pure inputs $\psi _{in}^{1}$ and $\psi _{in}^{2}$
respectively, with 
\begin{equation}
\nonumber
N_{i}^{\mu }\left( \varphi \right) =\sum\limits_{j=1,2}a_{ij}^{\mu }e^{%
i\kappa _{j}^{\mu }\varphi },\qquad \mu =1,2.
\end{equation}%
We can see that for these inputs the wave functions in the ring contain only
two of the four eigenstates of the Hamiltonian, those which have the same
spinor part.

By calculating the spin current densities corresponding to the $\Psi
_{ij}^{\mu }\left( \varphi \right) $ states appearing in (\ref{Psi_i_1_2})
we obtain%
\begin{equation}
J_{ij}^{\mu }=\left\vert a_{ij}^{\mu }\right\vert ^{2}\left[ 2\kappa
_{j}^{\mu }+\left( -1\right) ^{\mu }\left( \cos \theta -\frac{\omega }{%
\Omega }\sin \theta \right) \right] =\left( -1\right) ^{\mu
+j+1}2q\left\vert a_{ij}^{\mu }\right\vert ^{2}.  \label{J_ij_mu}
\end{equation}%
By examining (\ref{J_ij_mu}) we find that $\Psi _{i1}^{\mu }\left( \varphi
\right) $ and $\Psi _{i2}^{\mu }\left( \varphi \right) $\ represent
oppositely directed (clockwise and anticlockwise) spin currents in each
section (identified by the index $i$) of the ring, since $J_{i1}^{\mu }$ and 
$J_{i2}^{\mu }$\ have opposite signs. The overall spin current densities --
containing both clockwise and anticlockwise directed currents -- which
correspond to the input $\psi _{in}^{\mu }$ are%
\begin{eqnarray}
  J_{i}^{\mu }\! &=&\!2q\!\left( -1\right) ^{\mu }\!\left( \!\left\vert
a_{i1}^{\mu }\right\vert ^{2}\!-\!\left\vert a_{i2}^{\mu }\right\vert
^{2}\!\right) \!+\!2\mathrm{Re}\!\left( \!a_{i1}^{\mu }\left( a_{i2}^{\mu
}\right) ^{\ast }e^{i\left( \kappa _{1}^{\mu }-\kappa _{2}^{\mu
}\right) \varphi }\!\right) \left[ \!\kappa _{1}^{\mu }\!+\!\kappa _{2}^{\mu
}\!+\!\left( -1\right) ^{\mu }\!\left( \!\cos \theta \!-\!\frac{\omega }{%
\Omega }\sin \theta \!\right) \!\right]  \notag \\
  &=&\!2q\left( -1\right) ^{\mu }\!\left( \!\left\vert a_{i1}^{\mu
}\right\vert ^{2}\!-\!\left\vert a_{i2}^{\mu }\right\vert ^{2}\!\right)
=J_{i1}^{\mu }-J_{i2}^{\mu }.  \label{Jimu}
\end{eqnarray}%
We note that the disappearance of the cross terms in (\ref{Jimu}) is due to
the fact that $\tan \theta =-\omega /\Omega $.

The output spinors given by (\ref{out1}) and (\ref{out2}) suggest, that in
order to obtain a polarized (pure) state at a given output, we need one of
the one-dimensional projectors of (\ref{rho_i}) to vanish, and the other one
to remain nonzero at that point of the ring. In order to have different
polarized spin states in both outputs, the two projectors need to vanish at
the different output junctions.\ One of the possible ways to achieve this is
to have\ $\Psi _{I}^{1}\left( \gamma _{1}\right) $ and $\Psi _{II}^{1}\left(
\gamma _{2}\right) $ being zero, which happens if the spatial parts of these
wave functions are zero (see equations (\ref{Psi_i_1_2})): 
\begin{equation}
N_{I}^{1}\left( \gamma _{1}\right) =0,\qquad N_{II}^{2}\left( \gamma
_{2}\right) =0,  \label{cond1}
\end{equation}%
indicating destructive interference at the given output. By exchanging $\mu
\!=\!1$ and $\mu \!=\!2$, we can describe the other case of polarization.
Condition (\ref{cond1}) can be satisfied if%
\begin{equation}
\left\vert a_{I,1}^{1}\right\vert =\left\vert a_{I,2}^{1}\right\vert ,\qquad
\left\vert a_{I\!I,1}^{2}\right\vert =\left\vert a_{I\!I,2}^{2}\right\vert ,
\label{a_eq}
\end{equation}%
which, by using equation (\ref{a}), can be shown to be equivalent to
equations (\ref{detcond1}). (Exchanging $\mu \!=\!1$ and $\mu \!=\!2$ in (%
\ref{cond1}) leads to (\ref{detcond2})). Equation (\ref{a_eq}) implies that
the spin currents $J_{I}^{1}$ and $J_{I\!I}^{2}$\ given by (\ref{Jimu})
vanish as a consequence of the interference of oppositely directed currents
corresponding to states of the same spinor parts. We can see that the
requirement for spin-polarization given by (\ref{proj}), has a very clear
physical interpretation in terms of destructive interference and vanishing
spin currents. 
\begin{figure}[tbp]
\begin{center}
\includegraphics*[width=12 cm]{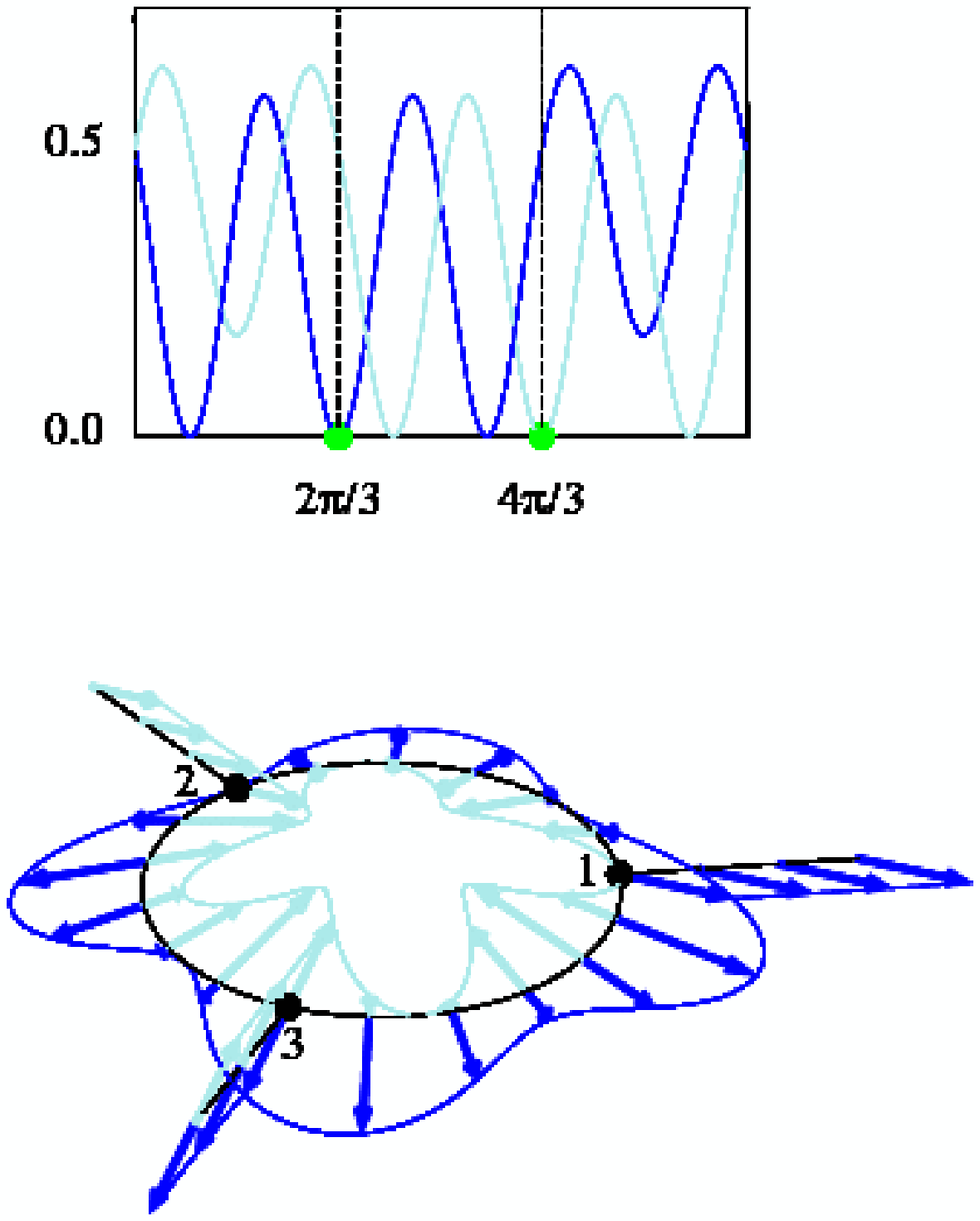}
\end{center}
\caption{The stationary spin directions of the electron along the
(symmetric) ring for a completely unpolarized input given by $\protect%
\varrho _{in}=\frac{1}{2}\left( \left\vert \protect\psi _{in}^{1}\right%
\rangle \left\langle \protect\psi _{in}^{1}\right\vert +\left\vert \protect%
\psi _{in}^{2}\right\rangle \left\langle \protect\psi _{in}^{2}\right\vert
\right) $ \ with $\protect\gamma _{2}=4\protect\pi /3$, $\protect\omega %
/\Omega =3.05$, $ka=1.38$ \protect\cite{note2}, which ensure perfect
polarization in the case given by equation (\protect\ref{detcond1}). Light
and dark (blue online) arrows correspond to $\protect\psi _{in}^{1}$ and $%
\protect\psi _{in}^{2}$ respectively. The length of the arrows as well as
the curves with the corresponding colour on the upper graph show the
probabilities of finding the electron at the given point on the ring. The
dashed lines in the upper graph mark the outgoing leads, where one of the
two probabilities becomes zero, resulting in the output of the other spinor
as a pure state. The two outputs in this case are given by equations (%
\protect\ref{out1}).}
\label{fig3}
\end{figure}

Figure \ref{fig3} and the corresponding animation show the stationary spin
directions of the electron along the ring for a completely unpolarized
input, for parameter values which ensure perfect polarization at the outputs
(in the case given by (\ref{detcond1}) for a symmetric ring). Light and dark
(blue online) arrows correspond to $\psi _{in}^{1}$ and $\psi _{in}^{2}$,
respectively. The length of the arrows as well as the curves with the
corresponding colour on the upper graph show the probabilities of finding
the electron at the given point on the ring. The dashed lines in the upper
graph mark the outgoing leads, where one of the two probabilities becomes
zero, leading to the output of the other spinor, given by equation (\ref%
{out1}). We note that spin transformation in this case is a rotation around
the z-axis by an angle pertaining to the given point on the ring.

\begin{figure}[tbp]
\begin{center}
\includegraphics*[width=12 cm]{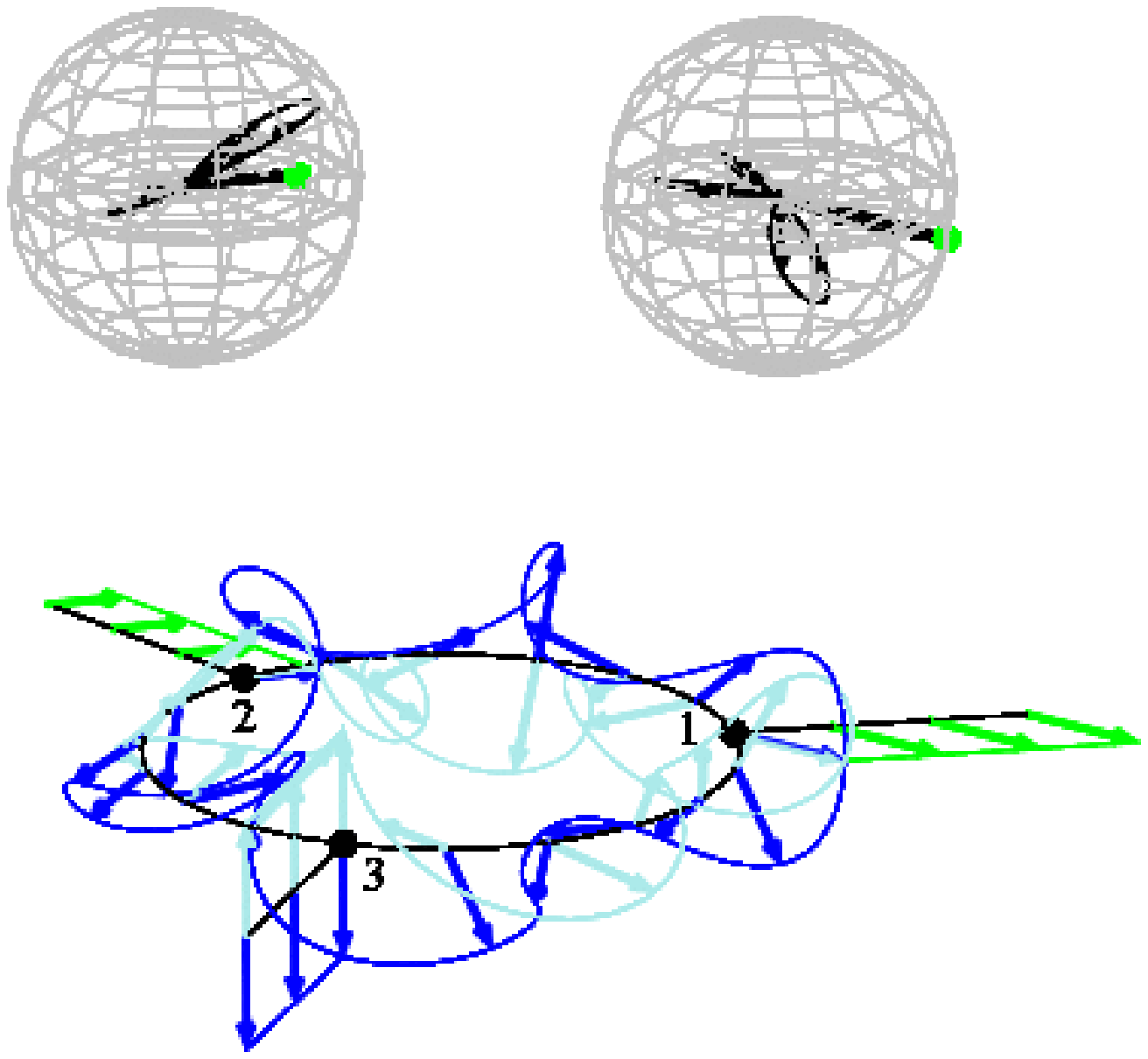}
\end{center}
\caption{The stationary spin directions along the ring for a completely
unpolarized input given by $\protect\varrho _{in}=\frac{1}{2}\left(
\left\vert \uparrow \right\rangle \left\langle \uparrow \right\vert
+\left\vert \downarrow \right\rangle \left\langle \downarrow \right\vert
\right) $, for the same parameter values as for figure \protect\ref{fig3}.
Light and dark (blue online) arrows correspond to inputs $\left\vert
\uparrow \right\rangle $\ and $\left\vert \downarrow \right\rangle $%
\thinspace\ respectively. The outgoing arrows (green online) represent the
output spinors given by equations (\protect\ref{out1}). The spin state of
the electron in left (right) branch of the ring is illustrated on the Bloch
spheres \protect\cite{NC00} above the respective part of the ring. At the
positions of the output junctions the black arrows reach the surface of the
spheres -- denoted by the dots (green online) -- indicating that the state
at those points is pure. The animation shows that the $\left\vert \uparrow
\right\rangle $ and $\left\vert \downarrow \right\rangle $ inputs are
rotated into the same direction at the outputs of the ring.}
\label{fig4}
\end{figure}

It is also interesting to see how polarization is produced if we decompose
the incoming perfect mixture as an equal weight sum of the eigenstates of $%
S_{z}$ 
\begin{equation}
\nonumber
\varrho _{in}=\frac{1}{2}\left( \left\vert \uparrow \right\rangle
\left\langle \uparrow \right\vert +\left\vert \downarrow \right\rangle
\left\langle \downarrow \right\vert \right) .
\end{equation}%
Figure \ref{fig4} and the corresponding animation show the stationary spin
directions along the ring for such an input and for the same parameter
values as in figure \ref{fig3}. Light and dark (blue online) arrows
correspond to inputs $\left\vert \uparrow \right\rangle $\ and $\left\vert
\downarrow \right\rangle $,\ respectively. The outgoing arrows (green
online) represent the output spinors given by equations (\ref{out1}). The
spin direction of the electron in the two branches of the ring is
illustrated on the two Bloch spheres \cite{NC00} above the ring, where the
length of the black arrow represents the purity of the given state. When the
arrow reaches the surface, the spin state is pure, otherwise it is mixed,
zero length meaning a perfect mixture. The dots on the Bloch spheres (green
online) indicate that at the positions of the output junctions the states
are pure. The animation shows that the $\left\vert \uparrow \right\rangle $
and $\left\vert \downarrow \right\rangle $ inputs are rotated into the same
direction at the outputs of the ring, resulting in the same pure states as
those seen in figure \ref{fig3}.

The role of spin-orbit interaction in the polarizing process can also be
seen in figures \ref{fig3} and \ref{fig4}. The spin-sensitivity of the
problem leads to a symmetry \cite{Y06} of the stationary solution that is
more complex than the pure geometrical mirror transformation. Note that
without spin-orbit interaction destructive interference for a given spin
direction would imply that the orthogonal spin component of the wave
function is also zero at that point. Consequently placing the output
junctions in such positions would mean zero transmission probability. This
can also be seen by considering that the rotation of the $\left\vert
\uparrow \right\rangle $ and $\left\vert \downarrow \right\rangle $\ spinors
shown in figure \ref{fig4} is due to spin-orbit interaction, for $\omega =0$
(or $\Phi =0$) their direction in the ring would not change, i.e., they
could never precess into the same direction.

Figure \ref{fig3}\ and \ref{fig4} show that besides the actual output
junctions, which are situated symmetric with respect to the incoming lead,
there are additional points on both branches of the ring where the state of
the electron is pure. These points are situated asymmetric with respect to
the incoming lead, and can be determined from equations (\ref{gamma1}) and (%
\ref{gamma2}). If the output leads are put into those positions, the
outcoming spins states are the ones given by (\ref{out1}).

\bigskip

\section{Conclusions}

We considered a three-terminal quantum ring with one input and two output
leads, which for appropriate parameter values acts as a spin
polarizer, similarly to the Stern-Gerlach apparatus. We presented a detailed
analytic solution of the spin-dependent transport problem and provided the
physical interpretation of the process: For both symmetric and non-symmetric
geometries, polarization is due to spatial interference. At a given junction
this interference is destructive for a certain spin direction, while
constructive for its orthogonal counterpart, which, consequently is
transmitted into the output lead.

\bigskip

\section*{Acknowledgement}

This work was supported by the Flemish-Hungarian Bilateral Programme, the
Flemish Science Foundation (FWO-Vl), the Belgian Science Policy and the
Hungarian Scientific Research Fund (OTKA) under Contracts Nos. T48888,
D46043, M36803, M045596. One of us (O. K.) was supported by an EU-Marie
Curie training fellowship.

\bigskip

\appendix

\section*{Appendix}

\setcounter{section}{1}

Considering the input junction, the continuity and Griffith conditions read:%
\begin{equation}
\Psi _{3}\left( 0\right) =\Psi _{I}\left( 0\right) =\Psi _{I\!I\!I}\left(
2\pi \right) ,  \label{cont}
\end{equation}%
\begin{equation}
J_{3}\left( 0\right) -J_{I}\left( 0\right) +J_{I\!I\!I}\left( 2\pi \right)
=0,  \label{curr}
\end{equation}%
respectively. Analogous equations can be written for the other two
junctions. The appropriately normalized spin current densities in the leads
are given by%
\begin{equation}
J_{l}\left( x_{l}\right) =2a\mathrm{Re}\left( \Psi _{l}^{\dagger }\left(
x_{l}\right) \left( -i\frac{\partial }{\partial x_{l}}\right) \Psi
_{l}\left( x_{l}\right) \right) ,  \label{Jm}
\end{equation}%
and in the ring by \cite{MPV04}%
\begin{equation}
J_{i}\left( \varphi \right) =2\mathrm{Re}\left[ \Psi _{i}^{\dagger }\left(
\varphi \right) \left( \frac{\omega }{2\Omega }\sigma _{r}\left( \varphi
\right) -i\frac{\partial }{\partial \varphi }\right) \Psi _{i}\left(
\varphi \right) \right] ,  \label{Ji}
\end{equation}%
where $\sigma _{r}\left( \varphi \right) =\cos \varphi \sigma _{x}+\sin
\varphi \sigma _{y}$, $i=I,I\!I,I\!I\!I$ and $l=1,2,3$. For the sake of
definiteness, we present the details for the incoming junction (\textbf{3}).
The results for the other junctions can be obtained in a similar manner.

First, we simplify equation (\ref{curr}) by\ using (\ref{Jm}) and (\ref{Ji})%
\begin{equation}
a\left. \frac{\partial \Psi _{3}}{\partial x_{3}}\right\vert
_{x_{3}=0}-\left. \frac{\partial \Psi _{I}}{\partial \varphi }\right\vert
_{\varphi =0}+\left. \frac{\partial \Psi _{I\!I\!I}}{\partial \varphi }%
\right\vert _{\varphi =2\pi }=0.  \label{curr2}
\end{equation}%
Substituting the wave functions (\ref{Psi_i}) and (\ref{waves}) into (\ref%
{cont}) we get%
\begin{eqnarray}
f_{\uparrow }+r_{\uparrow } &=&\sum\limits_{\mu ,j}a_{I,j}^{\mu }u_{j}^{\mu
}=-\sum\limits_{\mu ,j}a_{I\!I\!I,j}^{\mu }e^{i\kappa _{j}^{\mu }2\pi
}u_{j}^{\mu },  \label{A1.1} \\
f_{\downarrow }+r_{\downarrow } &=&\sum\limits_{\mu ,j}a_{I,j}^{\mu
}v_{j}^{\mu }=-\sum\limits_{\mu ,j}a_{I\!I\!I,j}^{\mu }e^{i\kappa
_{j}^{\mu }2\pi }v_{j}^{\mu }.
\end{eqnarray}%
The same substitution in (\ref{curr2}) yields%
\begin{eqnarray}
f_{\uparrow }-r_{\uparrow } &=&\sum\limits_{\mu ,j}\beta _{j}^{\mu }\left(
a_{I,j}^{\mu }+a_{I\!I\!I,j}^{\mu }e^{i\kappa _{j}^{\mu }2\pi }\right)
u_{j}^{\mu }, \\
f_{\downarrow }-r_{\downarrow } &=&\sum\limits_{\mu ,j}\gamma _{j}^{\mu
}\left( a_{I,j}^{\mu }+a_{I\!I\!I,j}^{\mu }e^{i\kappa _{j}^{\mu }2\pi
}\right) v_{j}^{\mu },  \label{B1.2}
\end{eqnarray}%
where $\beta _{j}^{\mu }=\left( \kappa _{j}^{\mu }-\frac{1}{2}\right) /ka$
and $\gamma _{j}^{\mu }=\left( \kappa _{j}^{\mu }+\frac{1}{2}\right) /ka$.
Since our aim is to determine the $a_{ij}^{\mu }$ coefficients, we eliminate 
$r_{\uparrow }$ and $r_{\downarrow }$ 
\begin{eqnarray}
\sum\limits_{\mu ,j}\left( a_{I,j}^{\mu }+a_{I\!I\!I,j}^{\mu }e^{i%
\kappa _{j}^{\mu }2\pi }\right) u_{j}^{\mu } &=&0,  \label{1_} \\
\sum\limits_{\mu ,j}\left( a_{I,j}^{\mu }+a_{I\!I\!I,j}^{\mu }e^{i%
\kappa _{j}^{\mu }2\pi }\right) v_{j}^{\mu } &=&0,  \label{2_}
\end{eqnarray}%
\begin{eqnarray}
\sum\limits_{\mu ,j}\left[ \beta _{j}^{\mu }\left( a_{I,j}^{\mu
}+a_{I\!I\!I,j}^{\mu }e^{i\kappa _{j}^{\mu }2\pi }\right)
+a_{I,j}^{\mu }\right] u_{j}^{\mu } &=&2f_{\uparrow },  \label{3_} \\
\sum\limits_{\mu ,j}\left[ \gamma _{j}^{\mu }\left( a_{I,j}^{\mu
}+a_{I\!I\!I,j}^{\mu }e^{i\kappa _{j}^{\mu }2\pi }\right)
+a_{I,j}^{\mu }\right] v_{j}^{\mu } &=&2f_{\downarrow }.  \label{4_}
\end{eqnarray}

After substituting the $u_{j}^{\mu }$ and $v_{j}^{\mu }$ spinor components
given by (\ref{u1v1}) and (\ref{u2v2}) into (\ref{1_})-(\ref{4_}) we find%
\begin{eqnarray}
\sum\limits_{j}\left[ \left( a_{I,j}^{1}+a_{I\!I\!I,j}^{1}e^{i\kappa
_{j}^{1}2\pi }\right) \cos \frac{\theta }{2}+\left(
a_{I,j}^{2}+a_{I\!I\!I,j}^{2}e^{i\kappa _{j}^{2}2\pi }\right) \sin 
\frac{\theta }{2}\right] &=&0,  \label{1__} \\
\sum\limits_{j}\left[ \left( a_{I,j}^{1}+a_{I\!I\!I,j}^{1}e^{i\kappa
_{j}^{1}2\pi }\right) \sin \frac{\theta }{2}-\left(
a_{I,j}^{2}+a_{I\!I\!I,j}^{2}e^{i\kappa _{j}^{2}2\pi }\right) \cos 
\frac{\theta }{2}\right] &=&0,  \label{2__}
\end{eqnarray}%
\begin{eqnarray}
 \sum\limits_{j}\left\{ \left[ \beta _{j}^{1}\!\left(
a_{I,j}^{1}\!+\!a_{I\!I\!I,j}^{1}e^{i\kappa _{j}^{1}2\pi }\right)
\!+\!a_{I,j}^{1}\right] \cos \frac{\theta }{2}\!+\!\left[ \beta
_{j}^{2}\!\left( a_{I,j}^{2}\!+\!a_{I\!I\!I,j}^{2}e^{i\kappa
_{j}^{2}2\pi }\right) \!+\!a_{I,j}^{2}\right] \sin \frac{\theta }{2}\right\}
\! &=&\!2f_{\uparrow },  \label{3__} \\
 \sum\limits_{j}\left\{ \left[ \gamma _{j}^{1}\!\left(
a_{I,j}^{1}\!+\!a_{I\!I\!I,j}^{1}e^{i\kappa _{j}^{1}2\pi }\right)
\!+\!a_{I,j}^{1}\right] \sin \frac{\theta }{2}\!-\!\left[ \gamma
_{j}^{2}\!\left( a_{I,j}^{2}\!+\!a_{I\!I\!I,j}^{2}e^{i\kappa
_{j}^{2}2\pi }\right) \!+\!a_{I,j}^{2}\right] \cos \frac{\theta }{2}\right\}
\! &=&\!2f_{\downarrow }.  \label{4__}
\end{eqnarray}%
\ Notice that certain terms can be cancelled out by using simple
trigonometric identities, giving:%
\begin{eqnarray}
\sum\limits_{j}\left( a_{I,j}^{1}+a_{I\!I\!I,j}^{1}e^{i\kappa
_{j}^{1}2\pi }\right) &=&0,  \label{1-} \\
\sum\limits_{j}\left( a_{I,j}^{2}+a_{I\!I\!I,j}^{2}e^{i\kappa
_{j}^{2}2\pi }\right) &=&0,  \label{2-}
\end{eqnarray}%
\begin{eqnarray}
 \sum\limits_{j}\left[ \left( 2\kappa _{j}^{1}-\cos \theta \right) \left(
a_{I,j}^{1}+a_{I\!I\!I,j}^{1}e^{i\kappa _{j}^{1}2\pi }\right)
+2kaa_{I,j}^{1}-\sin \theta \left( a_{I,j}^{2}+a_{I\!I\!I,j}^{2}e^{i%
\kappa _{j}^{2}2\pi }\right) \right] &=&2d^{1},  \label{3-} \\
 \sum\limits_{j}\left[ \left( 2\kappa _{j}^{2}+\cos \theta \right) \left(
a_{I,j}^{2}+a_{I\!I\!I,j}^{2}e^{i\kappa _{j}^{2}2\pi }\right)
+2kaa_{I,j}^{2}-\sin \theta \left( a_{I,j}^{1}+a_{I\!I\!I,j}^{1}e^{i%
\kappa _{j}^{1}2\pi }\right) \right] &=&2d^{2},  \label{4-}
\end{eqnarray}%
where 
\begin{eqnarray}
d^{1} &=&2ka\left( \cos \frac{\theta }{2}f_{\uparrow }+\sin \frac{\theta }{2}%
f_{\downarrow }\right) , \\
d^{2} &=&2ka\left( \sin \frac{\theta }{2}f_{\uparrow }-\cos \frac{\theta }{2}%
f_{\downarrow }\right) .
\end{eqnarray}%
The sums in equations (\ref{3-}) and (\ref{4-}) can be simplified using (\ref%
{1-}) and (\ref{2-}):%
\begin{eqnarray}
\sum\limits_{j}\left[ \left( 2\kappa _{j}^{1}+2ka\right) a_{I,j}^{1}+2\kappa
_{j}^{1}a_{I\!I\!I,j}^{1}e^{i\kappa _{j}^{1}2\pi }\right] &=&2d^{1},
\label{3} \\
\sum\limits_{j}\left[ \left( 2\kappa _{j}^{2}+2ka\right) a_{I,j}^{2}+2\kappa
_{j}^{2}a_{I\!I\!I,j}^{2}e^{i\kappa _{j}^{2}2\pi }\right] &=&2d^{2}.
\label{4}
\end{eqnarray}
Thus, the equations originating from the continuity requirements at the
incoming junction (\textbf{3}) (see figure \ref{fig1}) split into two
separate systems for $\mu =1,2$.

The other two junctions lead to four additional equations and consequently
we have to solve six equations for six unknowns for each $\mu $. We can
start by expressing $a_{I\!I\!I,j}^{1}$ by $a_{I,j}^{1}$ from equations (\ref%
{1-}) and (\ref{3-}), and also by $a_{I\!I,j}^{1}$ from the equations for
junction \textbf{2}. Using these two expressions for $a_{I\!I\!I,j}^{1}$ we
obtain a relation between $a_{I\!I,j}^{1}$ and $a_{I,j}^{1}$. From the
equations of the first outgoing junction we can also express $a_{I\!I,j}^{1}$
in terms of $a_{I,j}^{1}$. Finally, we can use the two different expressions
for $a_{I\!I,j}^{1}$ to calculate $a_{I,j}^{1}$. This automatically
determines all the other coefficients (since we have these expressed by $%
a_{I,j}^{1}$). The $\mu =2$ case can be solved analogously. The twelve $%
a_{ij}^{\mu }$ coefficients read

\begin{eqnarray}
  a_{I,j}^{\mu } &=&\frac{2d^{\mu }}{y}\left( -1\right) ^{\mu +j}\left\{
k^{2}a^{2}e^{i\left( -1\right) ^{\mu +j}q\gamma _{1}}\sin \left(
q\left( \gamma _{2}-\gamma _{1}\right) \right) \sin \left( q\left( 2\pi
-\gamma _{2}\right) \right) \right.   \notag \\
  &&\left. +i qka\left[ e^{i\left( -1\right) ^{\mu +j}q\gamma
_{1}}\sin \left( q\left( 2\pi -\gamma _{1}\right) \right) +e^{i\left(
-1\right) ^{\mu +j}q\gamma _{2}}\sin \left( q\left( 2\pi -\gamma _{2}\right)
\right) \right] \right.   \notag \\
  &&\left. -q^{2}\left( e^{i\left( -1\right) ^{\mu }w\pi }+e^{i%
\left( -1\right) ^{\mu +j}q2\pi }\right) \right\} ,  \notag \\
  a_{I\!I,j}^{\mu } &=&\frac{2qd^{\mu }}{y}\left( -1\right) ^{\mu
+j}\left\{ -q\left( e^{i\left( -1\right) ^{\mu }w\pi }+e^{i%
\left( -1\right) ^{\mu +j}q2\pi }\right) \right.   \notag \\
  &&\left. +i ka\left[ e^{i\left( -1\right) ^{\mu +j}q\gamma _{1}}%
e^{i\left( -1\right) ^{\mu }w\pi }\sin \left( q\gamma _{1}\right) +e%
^{i\left( -1\right) ^{\mu +j}q\gamma _{2}}\sin \left( q\left( 2\pi
-\gamma _{2}\right) \right) \right] \right\} ,  \notag \\
  a_{I\!I\!I,j}^{\mu } &=&\frac{1}{2q}\left[ \left( 2q+\left( -1\right)
^{\mu +j}ka\right) a_{I\!I,j}^{\mu }+\left( -1\right) ^{\mu +j}kae^{2i%
\left( -1\right) ^{\mu +j}q\gamma _{2}}a_{I\!I,j+\left( -1\right)
^{j+1}}^{\mu }\right] ,  \label{a}
\end{eqnarray}%
where 
\begin{eqnarray}
  y &=&i k^{3}a^{3}\left[ \sin \left( 2q\left( \pi -\gamma _{2}+\gamma
_{1}\right) \right) +\sin \left( 2q\left( \pi -\gamma _{1}\right) \right)
-\sin \left( 2q\left( \pi -\gamma _{2}\right) \right) -\sin \left( 2q\pi
\right) \right]   \label{y} \\
  &&-2qk^{2}a^{2}\left[ \cos \left( 2q\left( \pi -\gamma _{2}+\gamma
_{1}\right) \right) +\cos \left( 2q\left( \pi -\gamma _{1}\right) \right)
+\cos \left( 2q\left( \pi -\gamma _{2}\right) \right) -\cos \left( 2q\pi
\right) \right]   \notag \\
  &&+4qk^{2}a^{2}\cos \left( 2q\pi \right) -12i q^{2}ka\sin \left(
2q\pi \right) +8q^{3}\left[ \cos \left( w\pi \right) +\cos \left( 2q\pi
\right) \right] .  \notag
\end{eqnarray}

The elements of the transmission matrices can be determined from the
continuity equations $t_{1}^{\left( n\right) }=\sum\limits_{\mu
,j}a_{I\!I,j}^{\mu }e^{i\left( \kappa _{j}^{\mu }-\frac{1}{2}\right)
\gamma _{n}}u_{j}^{\mu }$ and $t_{2}^{\left( n\right) }=\sum\limits_{\mu
,j}a_{I\!I,j}^{\mu }e^{i\left( \kappa _{j}^{\mu }+\frac{1}{2}\right)
\gamma _{n}}v_{j}^{\mu }$ at the outgoing junctions (\textbf{1} and \textbf{2%
}) yielding (\ref{T_asymm}).

\bigskip

\providecommand{\newblock}{} \expandafter\ifx\csname url\endcsname\relax 

\fi \expandafter\ifx\csname urlprefix\endcsname\relax

\fi \providecommand{\eprint}[2][]{\url{#2}}


\begin{thebibliography}{99}
\bibitem{note2} We note that in semiconductor rings, the actual value of $ka$
is usually an order of magnitude larger, e.g. in a ring of radius $0.25\mu $%
m made of InGaAs, where the Fermi energy is $11.13$ meV, one has $%
k_{F}a\approx 20.4$. The experimentally feasible value of $\omega /\Omega $
is around $1$. The values used here are to provide a better visualization of
the phenomenon.
\end{thebibliography}
\end{document}